\documentclass[10pt]{article}

\usepackage{amsmath,amssymb,subfigure,comment}
\usepackage[algo2e,ruled,lined,titlenumbered]{algorithm2e}
\usepackage[ruled,section]{algorithm}
\newcommand{ \V }[1]{ \underline{#1} }
\newcommand{ \M }[1]{ \underline{\underline{#1}} }

\usepackage{geometry}
\usepackage{graphicx}
\usepackage{stmaryrd}
\usepackage{comment}

\title{A relocalization technique for the multiscale computation of delamination in composite structures.}

\author{O.~Allix, P.~Kerfriden, P.~Gosselet \vspace{5pt} \\ 
\large \textit{LMT-Cachan (ENS Cachan/CNRS/UPMC/PRES UniverSud Paris),} \\
\textit{61 av. du Pr\'esident Wilson, F-94230 Cachan, France}}

\date{February, 2010}

\begin{document}

\maketitle

\begin{abstract}

We present numerical enhancements of a multiscale domain decomposition strategy based on a LaTIn solver and dedicated to the computation of the debounding in laminated composites. We show that the classical scale separation is irrelevant in the process zones, which results in a drop in the convergence rate of the strategy. 
We show that performing nonlinear subresolutions in the vicinity of the front of the crack at each prediction stage of the iterative solver permits to restore the effectiveness of the method. \\

\end{abstract}

\section*{Introduction}

The reliable simulation of delamination in laminated composites requires fine models designed at the micro scale \cite{ladeveze06} (where fibers can be distinguished) or at most at the meso scale \cite{lubineau07,Cocchetti02} (where plies can be distinguished), because at these levels physics can be correctly taken into account. Thus, even the simulation of small laminated components implies to use huge discrete models which are out of the reach of non-optimized computational techniques. In this paper we mainly focus on two key ingredients to set up efficient strategies: the handling of nonlinearity and the use of nested scales of calculation to quickly distribute the computational effort on the whole structure --- or more precisely on all substructures since using domain decomposition is almost mandatory to achieve high-performance parallelism.

Most classical strategies provide separated answers to these two problems: nonlinearity is handled with Newton-Raphson algorithm (with, if required, arc-length control) \cite{LETALLEC:1994:NMN} or with asymptotic numerical method \cite{potierferry04}; multiscaling is realized on the linearized systems with domain decomposition methods like FETI \cite{farhat94}, BDD \cite{mandel93}, Schwarz \cite{lions90,ladeveze99,badea09}. Unfortunately, it is well known that the convergence of such approaches can be seriously impaired in the case of strong localized non-linearities \cite{cresta07}: not only the nonlinear process may require lots of load increments to converge but linear systems may be poorly conditioned and difficult to solve.

Recent studies have tried to take advantage of the domain decomposition to contain the difficulties associated to the nonlinearity within subdomains: they are based a process called nonlinear relocalization \cite{cresta07}  which consists in solving nonlinear problems independently inside subdomains (with well-chosen interface conditions). In \cite{pebrel08} this procedure was interpreted as conducting a Newton-Raphson on a nonlinear condensed problem, which enabled to classify variants and to propose new algorithms. Studies in \cite{gendre09} somehow consist in applying the relocalization philosophy under specific software constraints (use of a commercial ``closed'' finite element software).

Our studies are based on the LaTIn method which, from the very beginning \cite{ladeveze85}, has been designed as a nonlinear solver where nonlinearity was dealt with at the smallest possible scale (typically independently on each Gauss point for local material constitutive laws). The method was then adapted to substructuring by the introduction of unknown kinematic (displacement) and static (traction) interface fields which were linked by a constitutive equation (perfect joint, elastic joint, contact, friction). The Schwarz-type resolution algorithm was highly parallel, main operations were: resolution of nonlinear problem at the Gauss-point-scale, resolution of sparse linear systems at the subdomain-scale, exchange of interface vectors between subdomains \cite{ladeveze99}. Yet that local handling of information (subdomains communicating only with their neighbors) lead to a non-scalable algorithm: for a given problem, the convergence rate decreased when the number of subdomains increased. Scalability was achieved using a multiscale extension which consisted in insuring partial continuity and equilibrium conditions between subdomains which resulted in the resolution of a small global (defined on the whole structure) linear problem \cite{ladeveze00}. The weak continuity conditions (called ``macro'' conditions) were inspired from Saint-Venant principle and homogenization techniques \cite{sanchezpalencia74,feyel00}: the long-range influence of the physical phenomena was thus transmitted to the whole structure, so that subdomains not only got information from their neighbors but also from distant substructures. The number of iterations to converge thus became independent on the number of subdomains. Since then the method has been validated on various nonlinear problems \cite{nouy04,passieux09}.

Because of the presence of both kinematic and static fields on the interfaces, introducing cohesive behaviors \cite{allix98} on the interfaces is very easy in the LaTIn method \cite{kerfriden09}. Unfortunately for such interface behaviors, insuring the scalability of the method is not as straightforward: the stress singularity associated to the crack is not well captured within classical macro quantities which results in the long range effect not being transmitted and the convergence being seriously impaired (see also \cite{Jonsthovel09} for a somehow similar study). One interpretation of this phenomenon is that the very local treatment of nonlinearity prevent us from filtering the long range information to be transmitted globally. In this paper, we show that an intermediate treatment of the nonlinearity enables to detect the relevant information to spread along the structure. This technique is thus connected to the previously mentioned relocalization techniques, though here the treatment of nonlinearity is scaled-up instead of being scaled-down which seems more robust with respect to the risk of non-physical local instabilities, moreover very pertinent boundary conditions are easily imposed on the boundary of the relocalization domain. 

The rest of the paper is organized as follow:  in the first section, we present the classical LaTIn method; in the second section, we show that damaging cohesive behavior leads to the loss of the scalability of the method and that strategies based on the a priori enrichment of the macro information to transmit are doomed to inefficiency; the third section presents the relocalization technique; the fourth section opens the discussion on the method.

\section{Reference debounding problem and resolution strategy}
\label{sec:reference_problem}
\subsection{Substructuring of the debounding problem}


The laminated structure $\mathbf{E}$ occupying the domain $\Omega$ is made out of adjacent plies, separated by cohesive interfaces.  An external traction field $\V{F}_d$ (respectively displacement field $\V{U}_d$) is prescribed on Part $\partial \Omega_f$ (respectively $ \partial \Omega_u$) of the boundary $\partial \Omega$. The volume force is denoted $\V{f}_d$. The simulation is performed under the assumption of small perturbations and the evolution over time is supposed to be quasi-static and isothermal; classical incremental scheme is used.


\begin{figure}[ht]
       \centering
       \includegraphics[width=0.7 \linewidth]{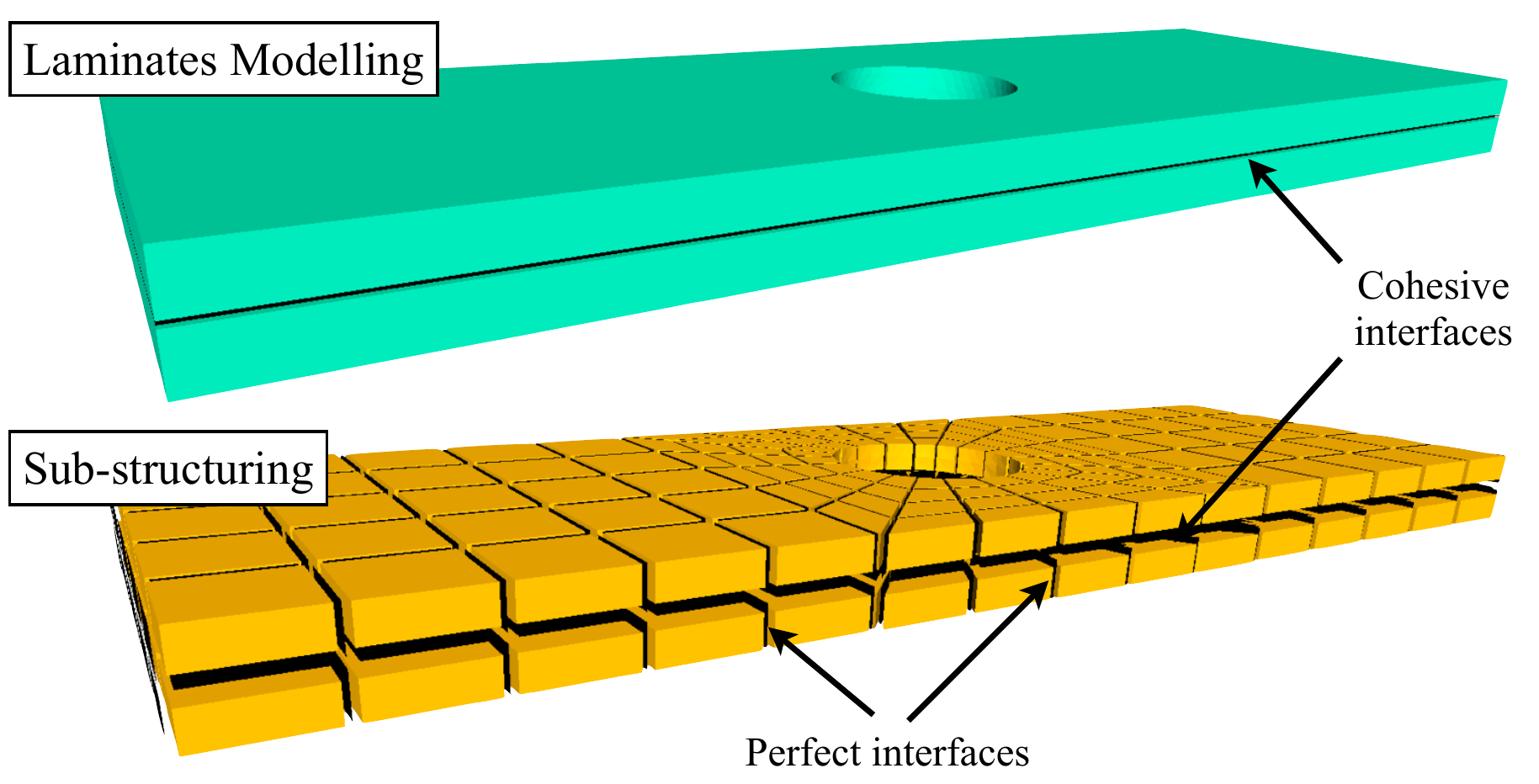}
       \caption{Substructuring of the composite structure}
       \label{fig:decomp_sst_interfaces}
\end{figure}

\begin{figure}[ht]
       \centering
       \includegraphics[width=0.6 \linewidth]{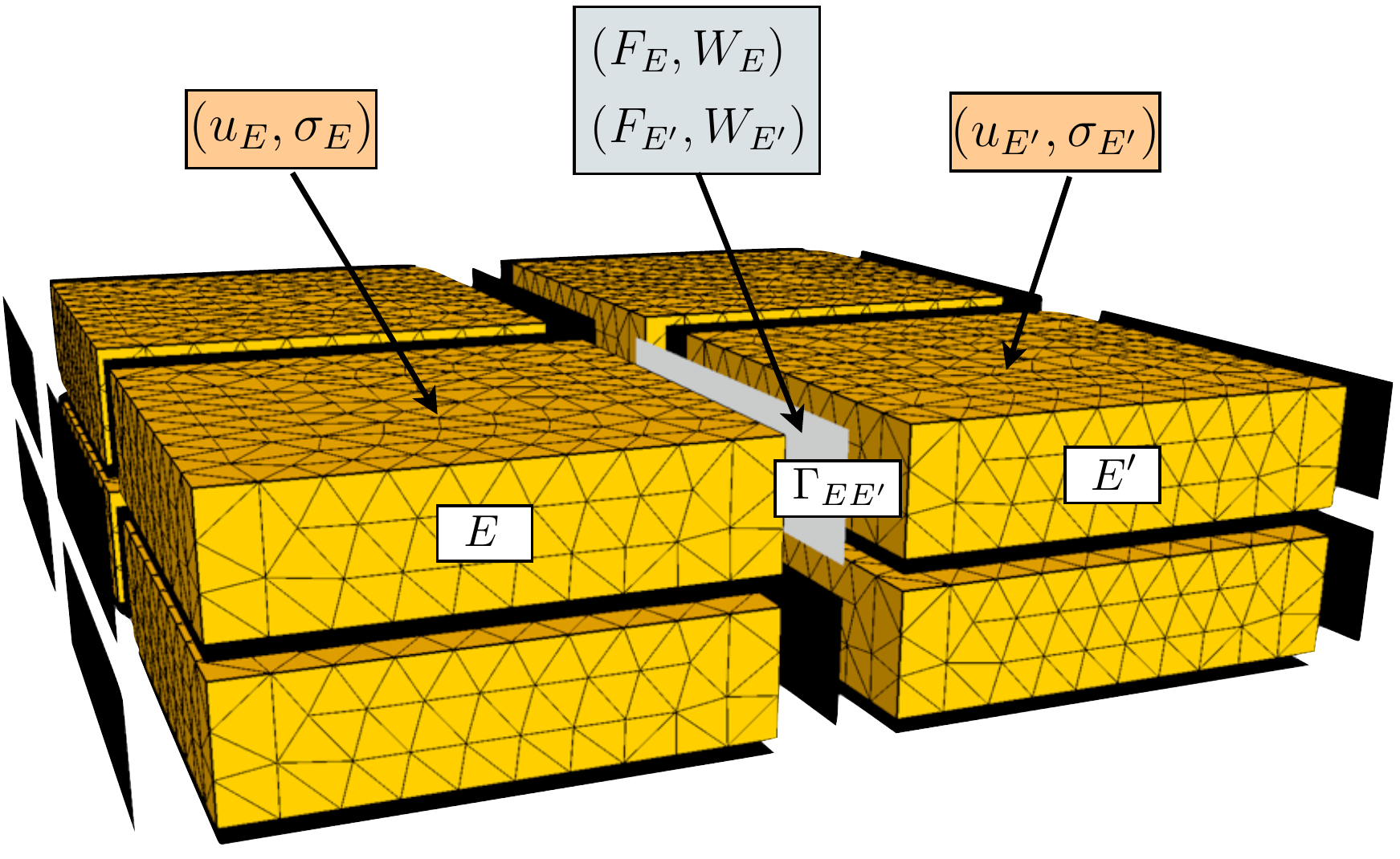}
       \caption{Mixed description of the unknown fields}
       \label{fig:champs_interface}
\end{figure}

The structure is decomposed into substructures and interfaces as represented Fig.~\ref{fig:decomp_sst_interfaces}. Each of these mechanical entities has its own kinematic and static unknown fields as well as its own constitutive law. The substructuring is driven by the will to match domain decomposition interfaces with material cohesive interfaces, so that each substructure $E$ belongs to a unique ply and has a constant linear constitutive law. Let $\M{\sigma}_E$ be the Cauchy stress tensor in $E$, $\V{u}_E$ the displacement field and $\M{\epsilon}(\V{u}_E)$ the symmetric part of the displacement gradient. A substructure $E$ defined in Domain $\Omega_E$ is connected to an adjacent substructure $E'$ through an interface $\Gamma_{EE'}=\partial \Omega_E \cap\partial \Omega_{E'}$ (Fig.~\ref{fig:champs_interface}). The surface entity $\Gamma_{EE'}$ applies force distributions $\V{F}_E$, $\V{F}_{E'}$ as well as displacement distributions $\V{W}_E$, $\V{W}_{E'}$ to $E$ and $E'$ respectively. Let us denote $\Gamma_E = \bigcup_{E' \in \mathbf{E}} \Gamma_{EE'} $. On a substructure $E$ such that $\partial \Omega_E \cap\partial \Omega\neq \emptyset$, the boundary condition $(\V{U}_d,\V{F}_d)$ is applied through a boundary interface $\Gamma_{E_d}$.

At each step of the incremental time resolution scheme, the substructured quasi-static problem consists in finding $s= (s_E)_{E \in \mathbf{E}} $, where $s_E = (\V{W}_E , \V{F}_E )$, which is a solution to the following equations:
\begin{itemize}
\item Kinematic admissibility of Substructure $E$:
\begin{equation}
\label{eq:kin_adm}
\textrm{at each point of } \Gamma_E, \quad {\V{u}_E} = {\V{W}_{E}}
\end{equation}
\item Static admissibility of Substructure $E$:
\begin{equation}
\label{eq:equilibre_sst}
 \begin{array}{l}
\displaystyle \forall ({\V{u}_E}^\star,{\V{W}_E}^\star) \in \mathcal{U}_{E} \times \mathcal{W}_{E} \  / \  {{\V{u}_E}^\star}_{| \partial \Omega_E} = {{\V{W}_{E}}^\star},  \\
\displaystyle \int_{\Omega_E} Tr \left( \M{\sigma}_E \, \M{\epsilon} ({\V{u}_E}^\star) \right) \ d \Omega =  \int_{\Omega_E} \V{f}_d . {\V{u}_E}^\star \, d \Omega+\displaystyle  \int_{\partial{\Gamma_E}} \V{F}_E . {\V{W}_E}^\star \ d \Gamma  
\end{array}
\end{equation}
\item Linear orthotropic constitutive law of Substructure $E$:
\begin{equation}
\label{eq:const_law_sst}
\textrm{at each point of } \Omega_E,\quad \M{\sigma}_E = \mathbf{K} \, \M{\epsilon}(\V{u}_E)
\end{equation}
\item Behavior of each interface $\Gamma_{EE'}$:
\begin{equation}
\label{eq:behavior_inter}
\displaystyle \textrm{at each point of } \Gamma_{EE'}, \quad
\mathcal{R}_{EE'}( \V{W}_{E}  , \V{W}_{E'} , \V{F}_{E} , \V{F}_{E'} ) = 0
\end{equation}
\item Behavior of the interfaces at the boundary $\partial \Omega_f \cap \Gamma_E$:
\begin{equation}
\label{eq:boundary_ref}
\begin{array}{l}
\textrm{at each point of } \Gamma_{{E}_d}, \quad \mathcal{R}_{E_d}( \V{W}_{E}, \V{F}_{E}) = 0\\
(\V{W}_{E}=\V{u}_d \  \text{ on } \partial\Omega_u \ \text{ and } \ \V{F}_{E}=\V{F}_d \ \text{ on }\partial\Omega_f)
\end{array}
\end{equation}
\end{itemize}

The formal relation $R_{EE'}=0$ is now made explicit in two representative cases:
  \begin{equation} \textrm{Perfect interface:} \qquad
\left\{ \begin{array}{l}
\V{F}_{E} + \V{F}_{E'} = 0 \\
\V{W}_{E} - \V{W}_{E'} = 0
\end{array} \right.
\end{equation}
\begin{equation} \textrm{Cohesive interface:} \qquad
\left\{ \begin{array}{l}
\displaystyle \V{F}_{E} + \V{F}_{E'} = 0 \\
\displaystyle \V{F}_{E} = \M{K}_{EE'} \left( (\V{[W]}_{EE' | \tau})_{\tau<t)} \right). \V{[W]}_{EE'} 
\end{array} \right.
\end{equation}
The last equation is the nonlinear constitutive law of the cohesive interfaces. The progressive softening of these interfaces (from healthy elastic to zero-stiffness in traction and shear) with respect to the history of the interface variables is described using continuum damage mechanics. Further details on the model used and on identification issues can be found in [\cite{allix98}].

\subsection{Two-scale iterative resolution of the substructured problem}
\label{sec:multiscale_resolution}

\subsubsection{Introduction of the macroscopic scale}

The substructured problem defined in previous section shall eventually be solved by an iterative LaTIn algorithm, which will be detailed in the next subsection. In order the strategy to be scalable (that is, for a given problem, to converge independently on the substructuring), a global coarse grid problem must be solved at each iteration of the solver. This coarse problem is associated to the equilibrium and continuity of so-called ``macroscopic'' force and displacement fields of the interfaces.

On each interface $\Gamma_{EE'}$ such that $(E,E') \in \mathbf{E}^2$, the interface fields are split into a macro part $.^M$ and a micro part $.^m$, the former belonging to a small-dimension subspace (9 macro degrees of freedom per
plane interface in 3D).
\begin{equation}
\displaystyle {\V{F}_E} = \V{F}_E^M + \V{F}_E^m \qquad \qquad
\displaystyle {\V{W}_E} = \V{W}_E^M + \V{W}_E^m
\end{equation}
The macro and micro data are uncoupled with respect to the interface virtual work:
\begin{equation}
\begin{array}{l}
\displaystyle \forall \displaystyle (\V{F}_E,\V{W}_E)\in\mathcal{F}_E\times\mathcal{W}_E, \quad 
\\ \displaystyle \int_{\Gamma_{EE'}} \V{F}_E.\V{W}_E \ d \Gamma 
=  \int_{\Gamma_{EE'}} \V{F}_E^M.\V{W}_E^M \ d \Gamma  
  + \int_{\Gamma_{EE'}} \V{F}_E^m.\V{W}_E^m \ d \Gamma
\end{array}
\end{equation}
Macro spaces are determined by the choice of their basis. Numerical tests have shown that using a linear macro basis provides, in general, good scalability properties to the method [\cite{ladeveze00}]. Indeed, the corresponding macro space includes the part of the interface fields with the highest wavelength. Consequently, according to the Saint-Venant principle, the micro complement found iteratively through the resolution of local problems only has a local influence.

\subsubsection{The iterative algorithm}

The iterative LaTIn algorithm [\cite{ladeveze99}], designed to solve nonlinear problems, is here applied to the resolution of the substructured debounding problem, the nonlinearities being lumped in the (cohesive) interfaces. 

The equations of the problem can be split into the set of linear equations in substructure and interface macroscopic variables (static and kinematic admissibility of the substructures, linear constitutive law of the substructures, linear equilibrium of the macro interface forces) and the set of local equations in interface variables (behavior of the interfaces). The solutions $\displaystyle s = (s_E)_{E \in \mathbf{E}} = (\V{W}_E ,  \V{F}_E )_{E \in \mathbf{E}}$ to the first set of equations belong to Space $\mathbf{A_d}$, while the solutions $\displaystyle \widehat{s} = (\widehat{s}_E)_{E \in \mathbf{E}} = (\V{\widehat{W}}_E ,  \V{\widehat{F}}_E )_{E \in \mathbf{E}}$ to the second set of equations belong to $\boldsymbol{\Gamma}$. Hence, the converged solution $s_{ref}$ is such that $ s_{ref} \in \mathbf{A_d} \bigcap \boldsymbol{\Gamma} $.

\begin{figure}[ht]
       \centering
       \includegraphics[width=0.5 \linewidth]{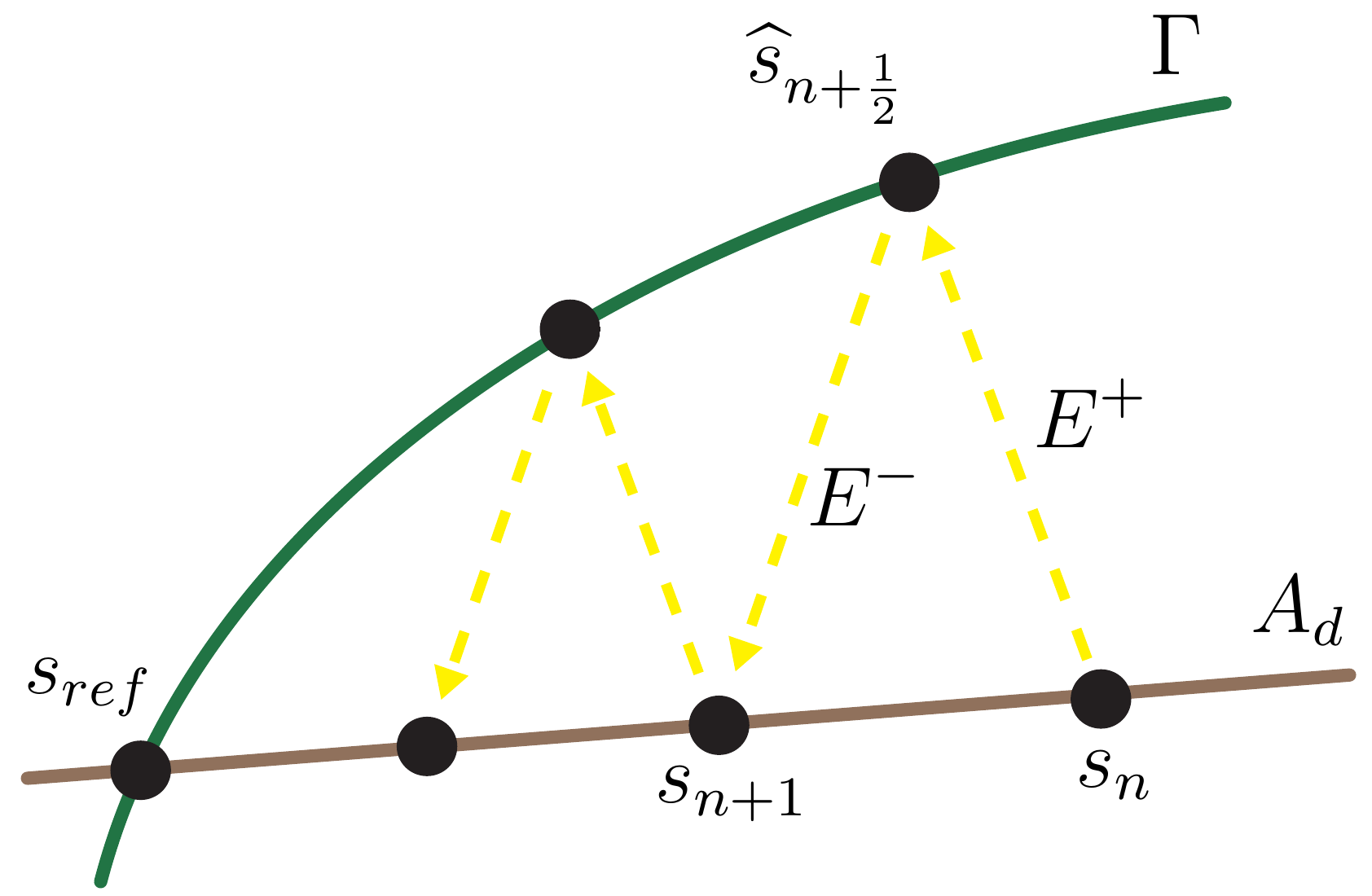}
       \caption{Illustration of the LaTIn iterative algorithm}
       \label{fig:latin}
\end{figure}

The resolution scheme consists in searching for the solution $s_{ref}$ 
alternatively in these two spaces along search directions  $\mathbf{E^+}$ and $\mathbf{E^-}$ (see Fig. \ref{fig:latin}): 
\begin{itemize}
\item 
\textit{Find}  $\widehat{s}_{n+\frac{1}{2}} \in \boldsymbol{\Gamma}  $ \textit{such that}  $ \left( \widehat{s}_{n+\frac{1}{2}} -s_{n} \right) \in \mathbf{E^+} $ (local stage)
\item 
\textit{Find} $s_{n+1} \in \mathbf{A_d}$ \textit{such that}  $\left( s_{n+1} - \widehat{s}_{n+\frac{1}{2}} \right) \in \mathbf{E^-} $ (linear stage)
\end{itemize}
In the following, the subscript $n$ will be dropped.

\paragraph{Local stage}

Independent local problems are solved at each point of the interfaces $(\Gamma_{EE'})_{|(E,E') \in \mathbf{E}^2}$ (and $(\Gamma_{E_d})_{E \in \mathbf{E}}$ for the interfaces belonging to $\partial \Omega_u \cup  \partial \Omega_f$):
\begin{equation}
\label{eq:local_problem}
\textit{Find} \ (\V{\widehat{F}}_E,\V{\widehat{W}}_E,\V{\widehat{F}}_{E'},\V{\widehat{W}}_{E'})  \vspace{0.2cm}
\textit{ such that: } \\
\left\{ \begin{array}{l}
\displaystyle \mathcal{R}_{EE'}(\V{\widehat{W}}_E,\V{\widehat{W}}_{E'},\V{\widehat{F}}_E,\V{\widehat{F}}_{E'}) = 0  \\
\displaystyle  (\V{\widehat{F}}_E-\V{F}_E) - k^+ (\V{\widehat{W}}_E - \V{W}_E) = 0 \\
\displaystyle  (\V{\widehat{F}}_{E'}-\V{F}_{E'}) - k^+ (\V{\widehat{W}}_{E'} - \V{W}_{E'}) = 0
\end{array} \right.
\end{equation}
The last two equations are the search direction equations $\mathbf{E}^+$. For a cohesive interface, Problem \eqref{eq:local_problem} is nonlinear, and solved by a Newton-Raphson scheme.


\paragraph{Linear stage}

The linear stage consists in solving linear systems in substructure variables under the constraint of macroscopic equilibrium of the interface forces:
\begin{equation}
\label{eq:macro_ad}
\begin{array}{l}
 \displaystyle \forall \ \Gamma_{EE'} \ / \ (E,E') \in \mathbf{E}^2,\quad \V{F}_E^M + \V{F}_{E'}^M =  0 \\
 \displaystyle \forall \ \Gamma_{E_d} \ / \ \Gamma_E \cap \partial \Omega_f \neq 0, \quad \V{F}_E^M - \V{F}_{d}^M = 0
 \end{array}
\end{equation}
The macroscopic condition is not compatible with the monoscale search direction $\mathbf{E}^-$ coupling the interface displacement and forces fields at the linear stage. Hence the search direction is weakened and verified at best under the macroscopic constraint [\cite{ladeveze03b}]. Technically this is realized using a Lagrangian whose stationarity leads to a modified local search direction:
\begin{equation} 
\label{eq:ddr_loc}
\displaystyle \forall  {\V{W}_E}^\star \in \mathcal{W}_E,  \quad \int_{\Gamma_E} \left( \V{F}_{E}-   \V{\widehat{F}}_{E} + k^- \ (\V{W}_{E}-\V{\widehat{W}}_{E})   - k^- \V{\widetilde{W}}^M \right) . {\V{W}_E}^\star \ d\Gamma= 0
\end{equation}
where the Lagrange multiplier $\V{\widetilde{W}}^{M}$ is a macroscopic unknown of Interface $\Gamma_{EE'}$.

The expression of the problem to solve on each substructure $E$ is obtained by substituting \eqref{eq:ddr_loc} in \eqref{eq:equilibre_sst}:
\begin{equation}
\label{eq:pb_micro}
\begin{array}{ll}
\displaystyle \forall ({\V{u}_E}^\star,{\V{W}_E}^\star) & \in \ \ \displaystyle {\mathcal{U}_E} \times {\mathcal{W}_E}, 
\\
& \displaystyle  \int_{\Omega_E} Tr (\M{\epsilon} (\V{u}_E) \,  K \M{\epsilon} ({\V{u}_E}^\star) ) \ d\Omega + \int_{\Gamma_E} k^- \, \V{W}_E . {\V{W}_E}^\star \ d \Gamma 
 \\
& \displaystyle =  \int_{\Omega_E} \V{f}_d . {\V{u}_E}^\star \, d \Omega + \int_{\Gamma_E} (\V{\widehat{F}}_E +k^- \V{\widehat{W}}_E +k^- {\V{\widetilde{W}}}^M) . {\V{W}_E}^\star \ d \Gamma
\end{array}
\end{equation}
The condensation of this equation on the macro degrees of freedom leads to a relation coupling $\V{F}_{E}^M$ and $\V{\widetilde{W}}_{E}^M$ which can be introduced in the macro equilibrium equation \eqref{eq:macro_ad}. Eventually, one gets a small linear system defined on macro degrees of freedom. All subdomains contribute to that ``global'' system through explicitly computed homogenized (condensed) flexibilities $\mathbb{L}_E^M$. 
\begin{equation}
\label{eq:pb_macro}
\begin{array}{ll}
\displaystyle \forall  \displaystyle \V {\widetilde{W}}^{M \star} \in {\mathcal{W}^M}^0, & \ \displaystyle
\sum_E \int_{\Gamma_{E}} \mathbb{L}_E^M \ \V{\widetilde{W}}^M . \V {\widetilde{W}}^{M\star} \ d\Gamma
 \\ & \quad \displaystyle =
 \sum_E \int_{\Gamma_{E} \cap \partial \Omega_f} \V{F}_d .\V{\widetilde{W}}^{M\star} \ d\Gamma
 -  \sum_E \int_{\Gamma_{E}}  \V{\widetilde{F}}_E .\V{\widetilde{W}}^{M\star} \ d\Gamma
\end{array}
\end{equation}
The right-hand side of Equation \eqref{eq:pb_macro} can be interpreted as a macroscopic static residual obtained from the computation of a single-scale linear stage. In order to derive this term, the problem \eqref{eq:pb_micro} must be solved independently on each substructure (the expression of the local macroscopic contributions $(\V{\widetilde{F}}_E)_{E \in \mathbf{E}}$ is given in [\cite{kerfriden09}]). The resolution of the macroscopic problem \eqref{eq:pb_macro} leads to the global knowledge of Lagrange multiplier $\V{\widetilde{W}}^M$, which is finally used as prescribed displacement to solve the substructure problems \eqref{eq:pb_micro}.

In order to perform the resolutions of \eqref{eq:pb_micro} in substructure variables, finite element method is used. Since the constitutive law of the substructures is linear, the stiffness operator of each substructure can be factorized once at the beginning of the calculation and reused without updating throughout the analysis, which gives high numerical performance to the method.

Algorithm~\ref{alg:LaTIn} sums up the iterative procedure described in this section.

\begin{algorithm2e}[ht]\caption{The two-scale domain decomposition solver}\label{alg:LaTIn}
Construction of the stiffness operator of each substructure \;
Computation of the macro homogenized operator $\mathbb{L}_E^M$ of each substructure \;
Global assembly of the macroscopic operator\;
Initialization $s_0 \in \boldsymbol{\Gamma}$\;
\For{$n=0,\ldots,N$}{%
  \textbf{Linear stage:} computation of $s_n \in \mathbf{A_d}$ \;
   $\quad$ Computation of the macro right-hand term $\widetilde{F}_E$ of each substructure \;
   $\quad$ Global assembly of the macroscopic right-hand term \;
   $\quad$ Resolution of the macroscopic problem \eqref{eq:pb_macro} \;
   $\quad$ Resolution of the microscopic problems \eqref{eq:pb_micro} \;
  \textbf{Local stage:} computation of $s_{n+\frac{1}{2}} \in \boldsymbol{\Gamma}$ \;
   $\quad$ Resolution of the local problems \eqref{eq:local_problem} on Interfaces $(\Gamma_{EE'})_{(E,E') \in \mathbf{E}^2}$\;
   $\quad$ Resolution of the local problems on Boundary interfaces $(\Gamma_{E_d})_{E \in \mathbf{E}}$ \;
  Computation of a global error indicator (distance separating $\mathbf{A_d}$ and $\Gamma$)
}
\end{algorithm2e}

\section{Loss of scalability in the case of crack propagation}
\label{sec:lose_extens}
Our first tests have outlined a numerical issue arising when simulating delamination propagation with the multiscale domain decomposition method. We observed a drop of the convergence rate which, as explained in subsection \ref{subsec:sca_caus} can be interpreted as a loss of scalability and for which classical solution presented in subsection \ref{subsec:sca_rem} are not efficient. Next section proposes a more realistic solution to this problem.


\subsection{Causes}\label{subsec:sca_caus}

\begin{figure}[htb]
       \centering
       \includegraphics[width=0.65 \linewidth]{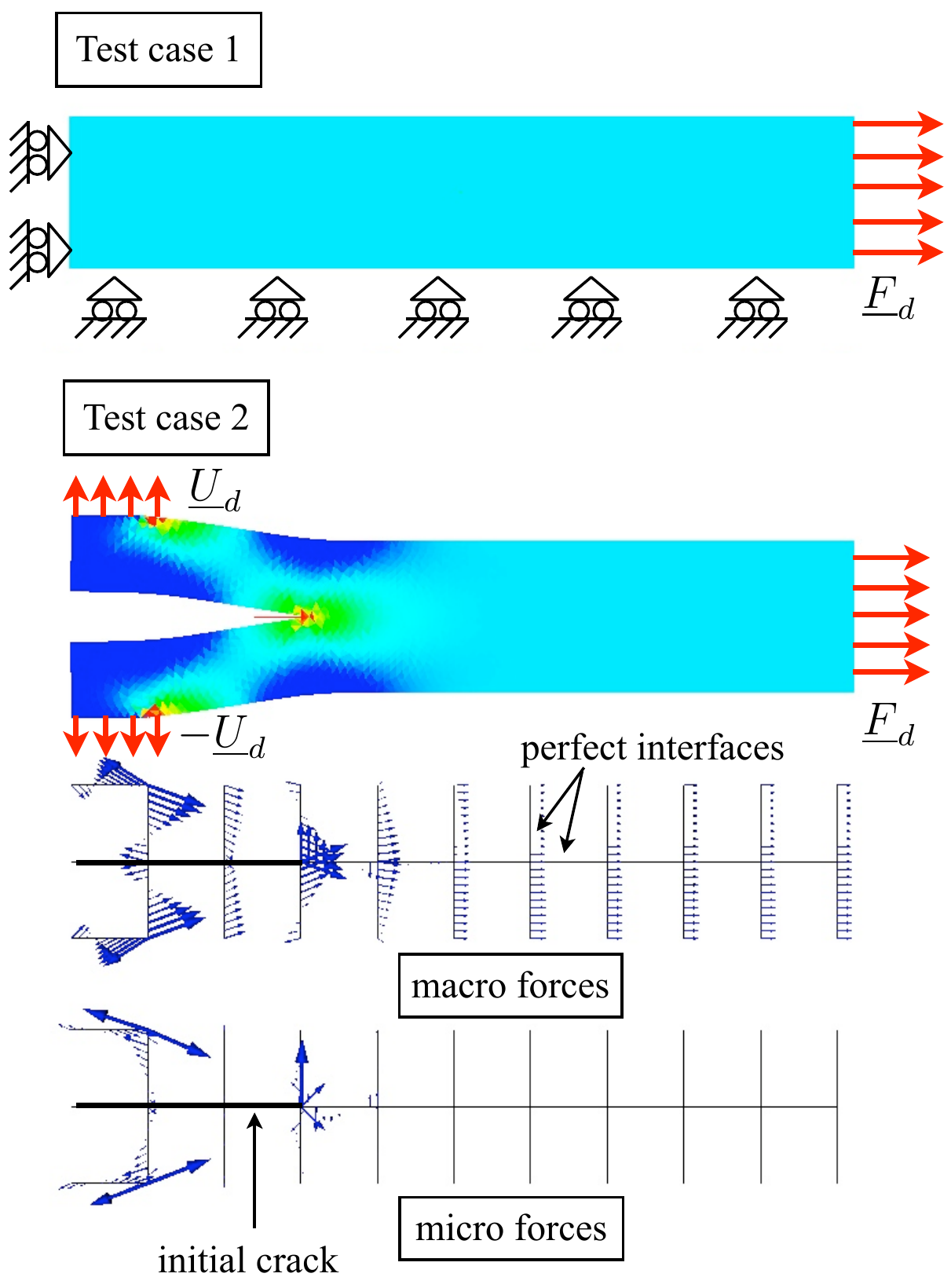}
       \caption{DCB-like 2D test cases. Macroscopic and microscopic solution forces are represented in Case 2.}
       \label{fig:test1_test2}
\end{figure}

\begin{figure}[htb]
       \centering
       \includegraphics[width=0.56 \linewidth]{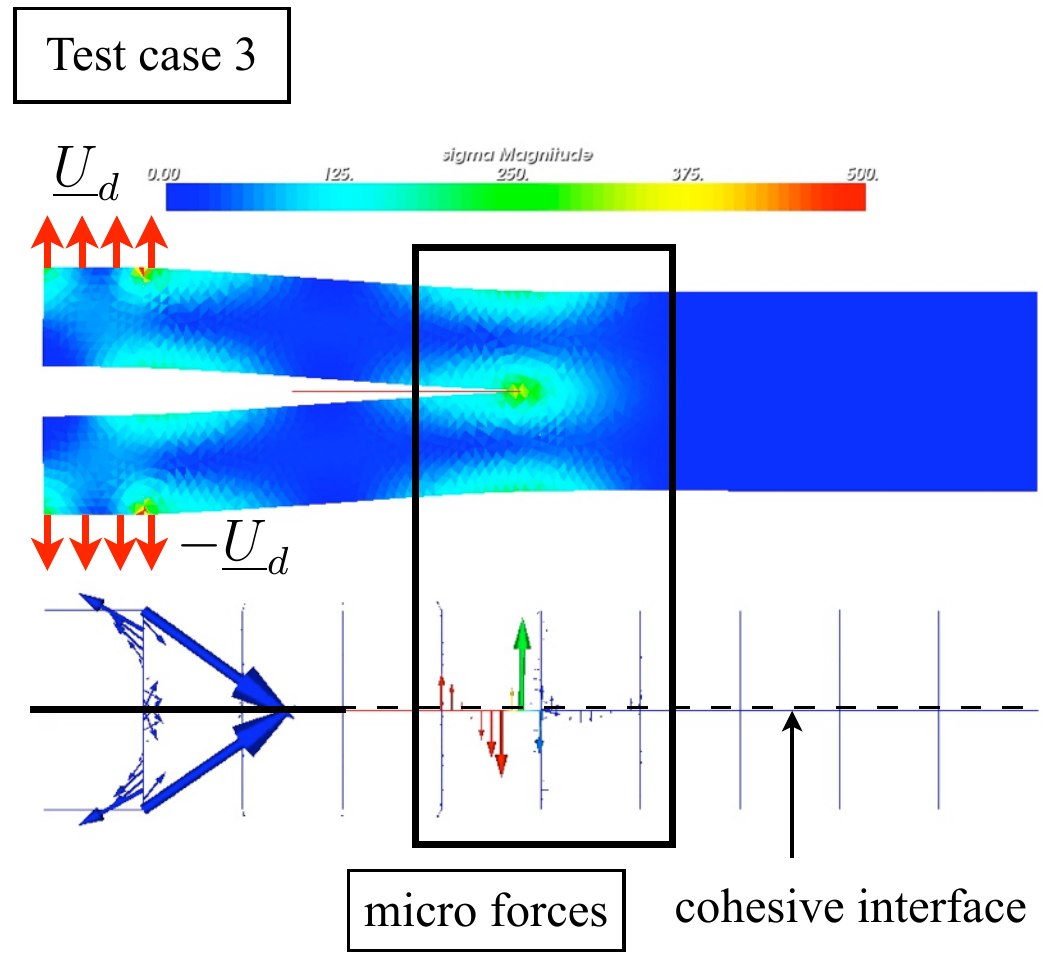}
       \caption{DCB 2D test case (cohesive interface separating isotropic plies). The microscopic stress concentration around the tip of the crack is not transmitted to the far substructures by solving the macroscopic problem.}
       \label{fig:test3}
\end{figure}

\begin{figure}[htb]
       \centering
       \includegraphics[width=0.6 \linewidth]{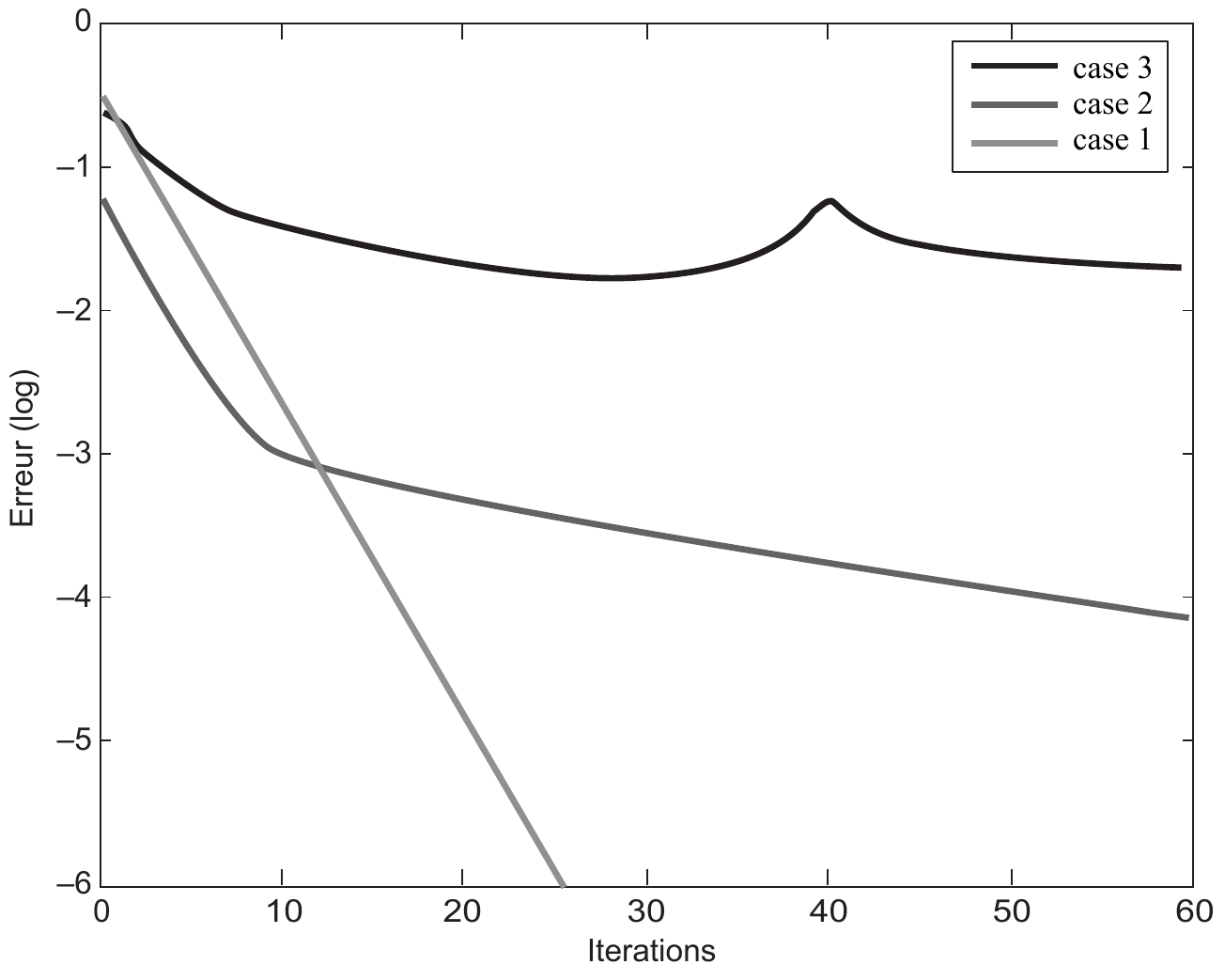}
       \caption{Convergence curve of the DCB-like test cases. The error criterion is a normalized distance separating Spaces $\mathbf{A_d}$ and $\boldsymbol{\Gamma}$ \cite{ladeveze99} }
       \label{fig:conv}
\end{figure}

Figure \ref{fig:test1_test2} shows the results of a computation performed on DCB-like (double cantilever beam) 2D test cases. The constitutive law of the plies is linear and isotropic, while the interfaces are perfect bonds, except three of them which simulate an pre-existing crack (unilateral frictionless contact). In Test case 1, the prescribed traction leads to the computation of a homogeneous solution. The fields are perfectly represented in the macroscopic space, which results in a high convergence rate of the strategy (Figure \ref{fig:conv}). In the second case, a stress singularity appears at the tip of the crack. Unfortunatly the classical scale separation is not adapted to that shape of stress field: a significant part of the stress field is orthogonal to the classical macroscopic space though its zone of influence is not confined (see the micro forces in Figure \ref{fig:test1_test2}). Hence local resolutions (with communications between adjancent substructures) are used to transmit large wavelenght information resulting in a drastic drop of the convergence rate of the LaTIn solver (Figure \ref{fig:conv}): in such a case the method is no more scalable.


In the last of these 2D test cases, we introduce the debounding ability by replacing some of the perfect bonds by cohesive interfaces (Figure \ref{fig:test3}). The immediate effect is a further drop in the convergence rate (Figure \ref{fig:conv}). Not only the stress singularity impairs the convergence (see micro forces distribution in Figure \ref{fig:test3}) but also, in order the tip of the crack to propagate from one Gauss point to its neighbor, a sufficiently converged global equilibrium state is indeed required. Hence, the solution to debounding problems requires the computation of successive quasi-equilibrated states within each load increment, which penalizes the numerical efficiency of the strategy.

\subsection{Remedies}\label{subsec:sca_rem}

Such a problem has already been encountered. In \cite{guidault08}, the authors simulate a crack propagation within the LaTIn-based multiscale framework, using the fracture mechanics theory. The cracks are, in this case, cutting the interfaces of the domain decomposition strategy, which results in a drop in the convergence rate. Scalability is successfully restored by enriching the macroscopic basis with piecewise linear functions in order the discontinuous interface displacement fields to be represented in the macroscopic space. 

\begin{figure}[htb]
       \centering
       \includegraphics[width=0.65 \linewidth]{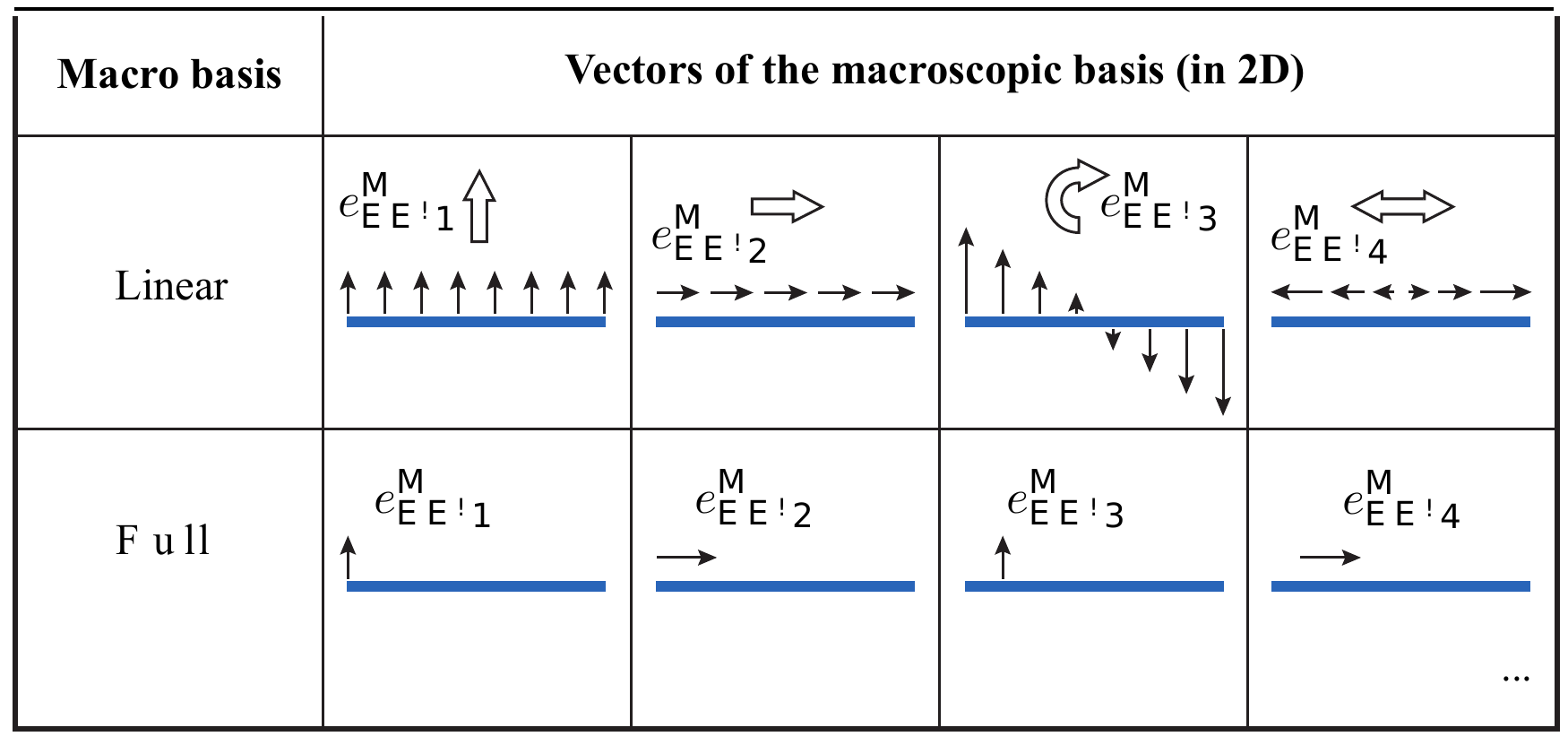}
       \caption{Linear macroscopic basis and finite element interface functions in 2D.}
       \label{fig:macro_basis}
\end{figure}

Following the same idea, we enrich the macroscopic basis in order to obtain a good representation of the stress concentration due to the cohesive crack in the macroscopic space. Though, in our case, the crack front position is not known in advance, and every potential position should be taken into account. Hence, no significant gain is obtained unless the macroscopic space is enriched locally around the tip of the crack (black square on Figure \ref{fig:test3}) by all the finite element interface functions (Figure \ref{fig:macro_basis}). This corresponds to clearing out the microspace and using maximal macrospace. 

\begin{figure}[htb]
       \centering
       \includegraphics[width=0.6 \linewidth]{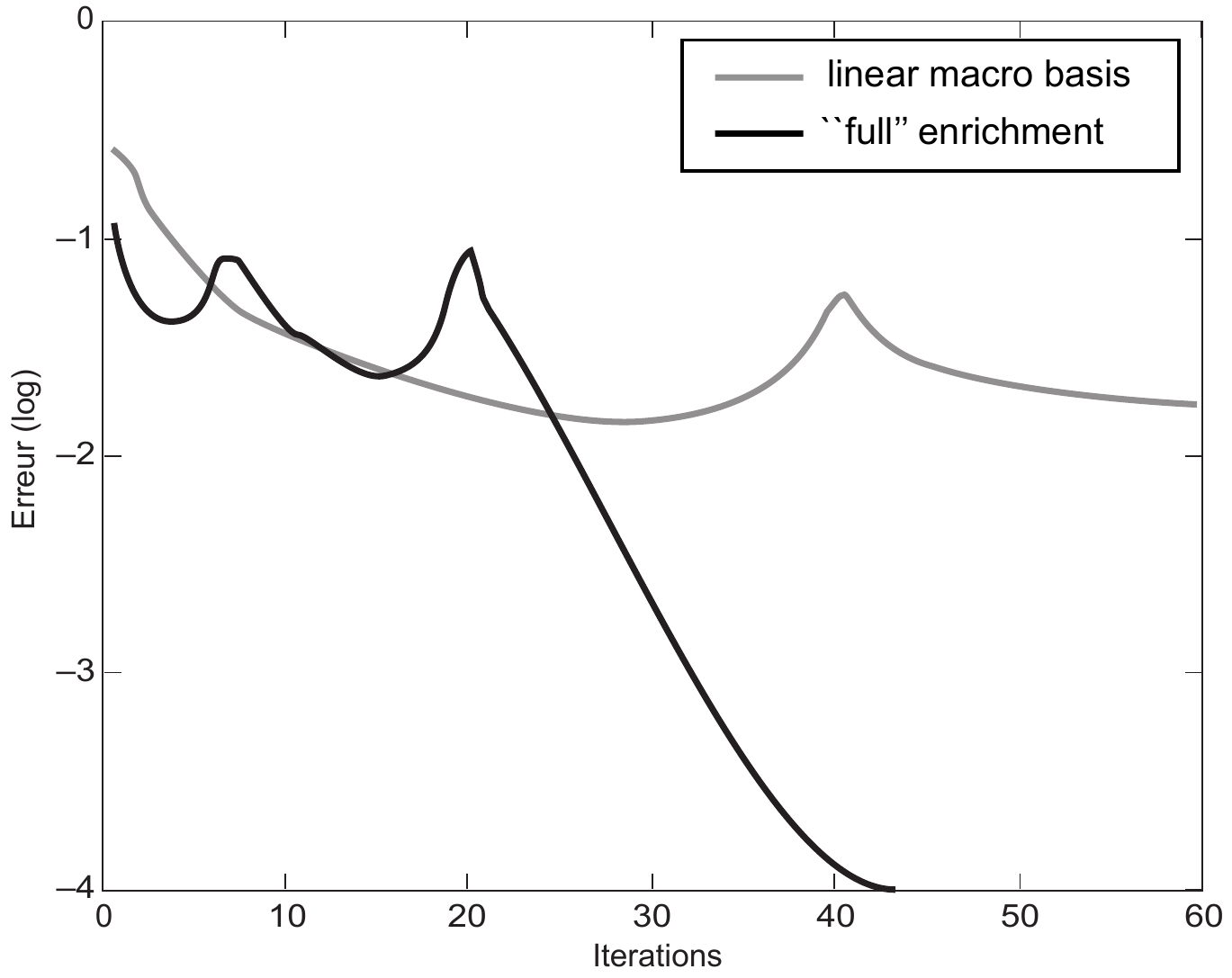}
       \caption{Convergence curve of the DCB test case using the locally enriched macroscopic basis.}
       \label{fig:conv2}
\end{figure}

The results obtained in terms of convergence rate are presented on Figure \ref{fig:conv2}. During the considered increment, the propagation spreads along the distance covered by three Gauss points; the damage state of one Gauss point needs to be sufficiently well converged before the next one starts damaging (which justifies the swaying shape of the curves), once the damage state of Gauss points is correctly acquiered then the whole structure converges (fast decreasing part of the black curve). The bump in the gray curve (classical macrospace) observed around Iteration 40 corresponds to the one observed around Iteration 8 of the black curve (full macrospace).

Though this enrichment is performed locally in small process zones, the extra computational costs are not affordable in 3D simulations. Indeed, computing the local homogenized operators $(\mathbb{L}_E^M)_{E \in \mathbf{E}}$ requires to perform a number of resolution of Problems \eqref{eq:pb_micro} equal to the number of interface degrees of freedom (explicit computation of Schur operators). 

Moreover, one can easily show \cite{kerfriden09} that such an enrichment, coupled with an adequate choice of the search direction $\mathbf{E}^-$ is equivalent to bond the subdomains with a linearized interface behavior equation \eqref{eq:behavior_inter} at the linear stages of the LaTIn solver. Hence, at this stage, a fully equilibrated solution is computed in the enriched zone.

The remedy that we proposed in this section thus implies lots of computations which makes it only adapted to 2D problems. Indeed in 3D, the number of degrees of freedom within the damaging zone is too large to consider full condensation. Then in the next section we propose an alternative strategy, suited to 3D problems, to compute this local solution.

\section{Relocalization strategy in the vicinity of the crack}
\label{sec:subiterations}
\begin{figure*}
       \centering
       \includegraphics[width=0.99 \linewidth]{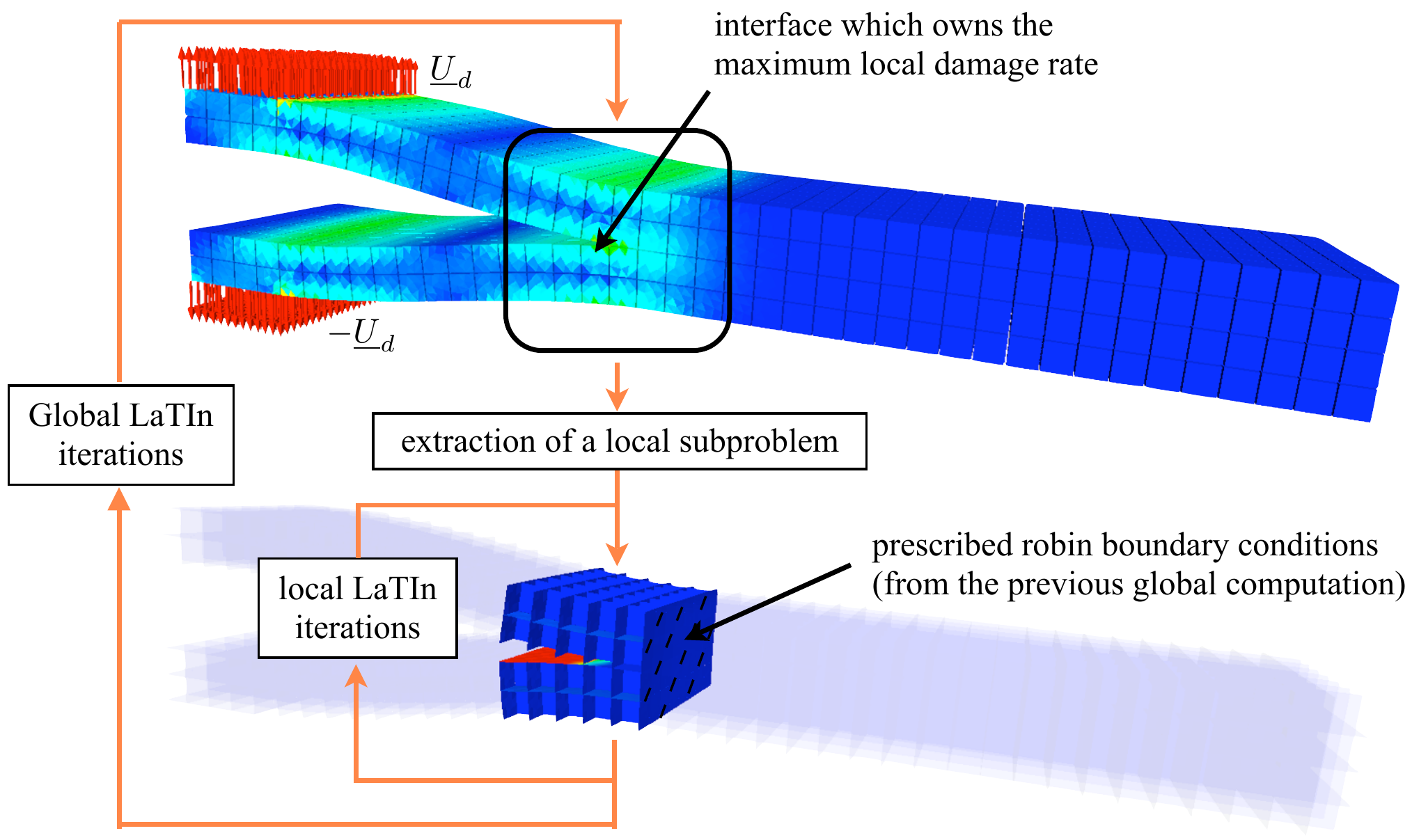}
       \caption{Illustration of the subresolution technique}
       \label{fig:adap_endom}
\end{figure*}

In order to illustrate the relocalization technique, we consider the 3D version of the DCB problem previously studied (see Figure \ref{fig:adap_endom}). The loading is applied in 10 increments, damage begins at the third one. As can be seem on Figure \ref{fig:sub_iterations} (``No subresolution'' data), as soon as damage appears the number of iterations per increment explodes (about 10 times more iterations). For such a problem, using a full macro basis would imply unrealistic extra computations.

\subsection{Principle of the sub-resolution strategy}

The alternative technique consists in extracting a part $\Omega_{sub}$ of Domain $\Omega$, where $\Omega_{sub}$ is a source of localized nonlinearities in the structure. At each linear stage of the global LaTIn solver, the converged solution of the nonlinear subproblem in $\Omega_{sub}$ is sought using the two-scale domain decomposition strategy described in Section \ref{sec:reference_problem} (see Figure (\ref{fig:adap_endom})) as together with Algorithm (\ref{alg:sub_iter})). 


\subsection{Local enhancement of the ``linear'' stage}


The new problem to solve at the enhanced ``linear'' stage now reads: \it Find $s= (s_E)_{E \in \mathbf{E}} $, \textit{where} $s_E = (\V{W}_E , \V{F}_E )$  solution to: \rm
\begin{itemize}
    \item linear equations in substructures variables in Domain $\Omega$
    \begin{itemize}
        \item static and kinematic admissibility of the substructures, Equations \eqref{eq:kin_adm} and \eqref{eq:equilibre_sst}
        \item linear constitutive law of the substructures, Equation \eqref{eq:const_law_sst}
    \end{itemize}
    \item interface bonding equations in $\Omega \backslash \Omega_{sub}$
    \begin{itemize}
     \item search direction $\mathbf{E}^-$
     \begin{equation}   
       \label{eq:search_sub}
            \forall \ \Gamma_{EE'} / (E \notin \Omega_{sub} \ \textrm{or}  \ E' \notin \Omega_{sub}), \quad \left(s-\widehat{s} \right) \in \mathbf{E}^-
     \end{equation}
     \item equilibrium of the macroscopic interface forces
             \begin{equation}
\label{eq:macro_sub}
 \forall \ \Gamma_{EE'} / (E \notin \Omega_{sub} \ \textrm{or}  \ E' \notin \Omega_{sub}), \quad \V{F}_E^M + \V{F}_{E'}^M =  0
\end{equation}
 \end{itemize}
     \item interface bonding equations in $\Omega_{sub}$: interface linear and nonlinear behavior 
     \begin{equation} \forall \ \Gamma_{EE'} / (E,E') \in \Omega_{sub}^2, \quad \mathcal{R}_{EE'}( \V{W}_{E}  , \V{W}_{E'} , \V{F}_{E} , \V{F}_{E'} )= 0  
     \end{equation}
          \begin{equation}
 \forall \  \Gamma_{{E}_d} / E \in \Omega_{sub} \ \textrm{and} \ \Gamma_E \cap(\partial \Omega_u \cap \partial \Omega_f)\neq0, \quad R_{Ed}( \V{W}_{E}, \V{F}_{E}) = 0
 \end{equation}
\end{itemize}

Solving these equations requires to perform a classical linear stage on the whole structure (as described in Sect. \ref{sec:reference_problem}), and a nonlinear ``exact'' subresolution on $\Omega_{sub}$.

\begin{algorithm2e}[ht]\caption{The subresolution strategy algorithm}\label{alg:sub_iter}
Construction of the stiffness operator of each substructure \;
Computation of the macro homogenized operator $\mathbb{L}_E^M$ of each substructure \;
Global assembly of the macroscopic operator\;
Initialization $s_0 \in \boldsymbol{\Gamma}$\;
\For{$n=0,\ldots,N$}{%
  \textbf{Enhanced linear stage:} computation of $s_n \in \mathbf{A_d}$ \;
  $\quad$ \textbf{Classical linear stage} on $\mathbf{E}$ \;
  $\quad$ \textbf{Subresolutions :} \;
  $\qquad$ Locate the process zones requiring subresolution \;
  $\qquad$ Assemble the macroscopic subproblem \;
  \For{$j=0,\ldots,m$}{ %
    Subproblem local stage  \;
    Subproblem linear stage  \;
    Computation of a local error indicator  \;
  }
  \textbf{Local stage:} computation of $s_{n+\frac{1}{2}} \in \boldsymbol{\Gamma}$ \;
  Computation of a global error indicator
}
\end{algorithm2e}

\subsection{Nonlinear subresolutions}

On the boundary $\partial\Omega_{sub} \backslash \partial \Omega$, the subproblem is bonded to the solution computed at the previous global linear stage. In order to enforce equations \eqref{eq:search_sub} and \eqref{eq:macro_sub}, macroscopic equilibrium and microscopic Robin boundary conditions are enforced:
\begin{equation}
\label{eq:bonding_sub}
\begin{array}{l}
 \forall \ \Gamma_{EE'} / (E \in \Omega_{sub} \ \textrm{and}  \ E' \notin \Omega_{sub}), \quad \\
\quad \left\{ \begin{array}{l}
 \displaystyle \forall \V{W}^{M \star} \in {\mathcal{W}^M}, \  \int_{\Gamma_{EE'}} \left( \V{F}_E - {\V{F}_E}_G \right).\V{W}^{M \star} \ d \Gamma = 0 \\
 \displaystyle \forall \V{W}^{m \star} \in {\mathcal{W}^m}, \ \int_{\Gamma_{EE'}} \left( \V{F}_E + k^- \V{W}_E - ({\V{F}_E}_G + k^- {\V{W}_E}_G)  \right) .\V{W}^{m\star} \ d \Gamma = 0
 \end{array} \right.
 \end{array}
\end{equation}
where ${\underline{F}_E}_G$ and ${\V{W}_E}_G$ are respectively the force and displacement fields $\underline{F}_E$ and $\underline{W}_E$ obtained at the end global linear stage on $(\Gamma_{EE'})_{(E \in \Omega_{sub} \ \textrm{and}  \ E' \notin \Omega_{sub})}$.

The set of equations in $\Omega_{sub}$ (substructures admissibility and constitutive law in $\Omega_{sub}$, interface behavior on $\Omega_{sub} \backslash \partial \Omega_{sub}$, boundary conditions on $\partial \Omega_{sub} \cap \partial \Omega$, bonding equations \eqref{eq:bonding_sub} on $\partial\Omega_{sub} \backslash \partial \Omega$) is solved using the LaTIn two-scale procedure described in Section 1. Two particular technical points should be focused on:
\begin{itemize}
\item The bonding equations \eqref{eq:bonding_sub} are handled at the local stage of this scheme in the same way the boundary equations \eqref{eq:boundary_ref} are dealt with in the classical two-scale resolution \cite{ladeveze99}. However, these two equations, coupled with the search direction $\mathbf{E}^+$, lead to the resolution of two global systems on each interface ${\Gamma_{EE'}}_{|(E \in \Omega_{sub} \ \textrm{and}  \ E' \notin \Omega_{sub})}$, for projections in the microscopic and macroscopic spaces are required.
\item At the linear stage of the subresolution scheme, the macroscopic equilibrium of $\Omega_{sub}$ in enforced (so as to ensure the scalability of the subresolution strategy):
\begin{equation}
\label{eq:macro_sub2}
\left\{ \begin{array}{l}
\forall \ \Gamma_{EE'} / (E \in \Omega_{sub} \ \textrm{or}  \ E' \in \Omega_{sub}), \quad \V{F}_E^M + \V{F}_{E'}^M =  0 \\
\forall \ \Gamma_{EE'} / (E \in \Omega_{sub} \ \textrm{and}  \ E' \notin \Omega_{sub}), \quad \V{F}_E^M - \V{F}_{EG}^M = 0 \\
\forall \ \Gamma_{E_d} / (E \in \ \Omega_{sub} \ \textrm{and} \ \Gamma_E \cap \partial \Omega_f \neq 0), \quad \V{F}_E^M - \V{F}_{d}^M = 0
\end{array} \right.
\end{equation}
The resulting macroscopic subproblem reads:
\begin{equation}
\label{eq:pb_macro_sub}
\begin{array}{l}
\displaystyle \forall  \V {\widetilde{W}}^{M \star} \in {\mathcal{W}_{sub}^M}^0, \\
 \displaystyle  \quad
\sum_{E \in \Omega_{sub}} \int_{\Gamma_{E}} \mathbb{L}_E^M \ \V{\widetilde{W}}^M . \V {\widetilde{W}}^{M\star} \ d\Gamma
 =
 \sum_{E \in \Omega_{sub}} \int_{\Gamma_{E} \cap ( \partial \Omega_{sub} \backslash \partial \Omega)} \V{F}_{E_G} .\V{\widetilde{W}}^{M\star} \ d\Gamma  \\ 
 \qquad \quad \displaystyle + \sum_{E \in \Omega_{sub}} \int_{\Gamma_{E} \cap \partial \Omega_f} \V{F}_d .\V{\widetilde{W}}^{M\star} \ d\Gamma
 -  \sum_{E \in \Omega_{sub}} \int_{\Gamma_{E}}  \V{\widetilde{F}}_E .\V{\widetilde{W}}^{M\star} \ d\Gamma
\end{array}
\end{equation}
The construction of this problem is cheap as the homogenized operators $\mathbb{L}_E^M$ are the same used for the global macroscopic problem (no enrichment of the macroscopic basis). Though, if $\textrm{mes} (\partial \Omega_{sub} \cap \partial \Omega_{u} = 0)$, the macroscopic assembled operator has a Kernel (rigid body motions of the structure extracted to perform the subresolution). Hence, Problem \eqref{eq:pb_macro_sub} must be solved using a generalized inverse.
\end{itemize}




\subsection{Results}
The example being treated is the DCB case represented Figure \ref{fig:adap_endom}. In order to extract the subproblem automatically, we choose to perform the subresolution on a set substructures and interfaces in a box surrounding the cohesive interface with the highest damage rate at the end of the previous global LaTIn iteration (see Figure (\ref{fig:adap_endom})).

\begin{figure}[htb]
       \centering
       \includegraphics[width=0.65 \linewidth]{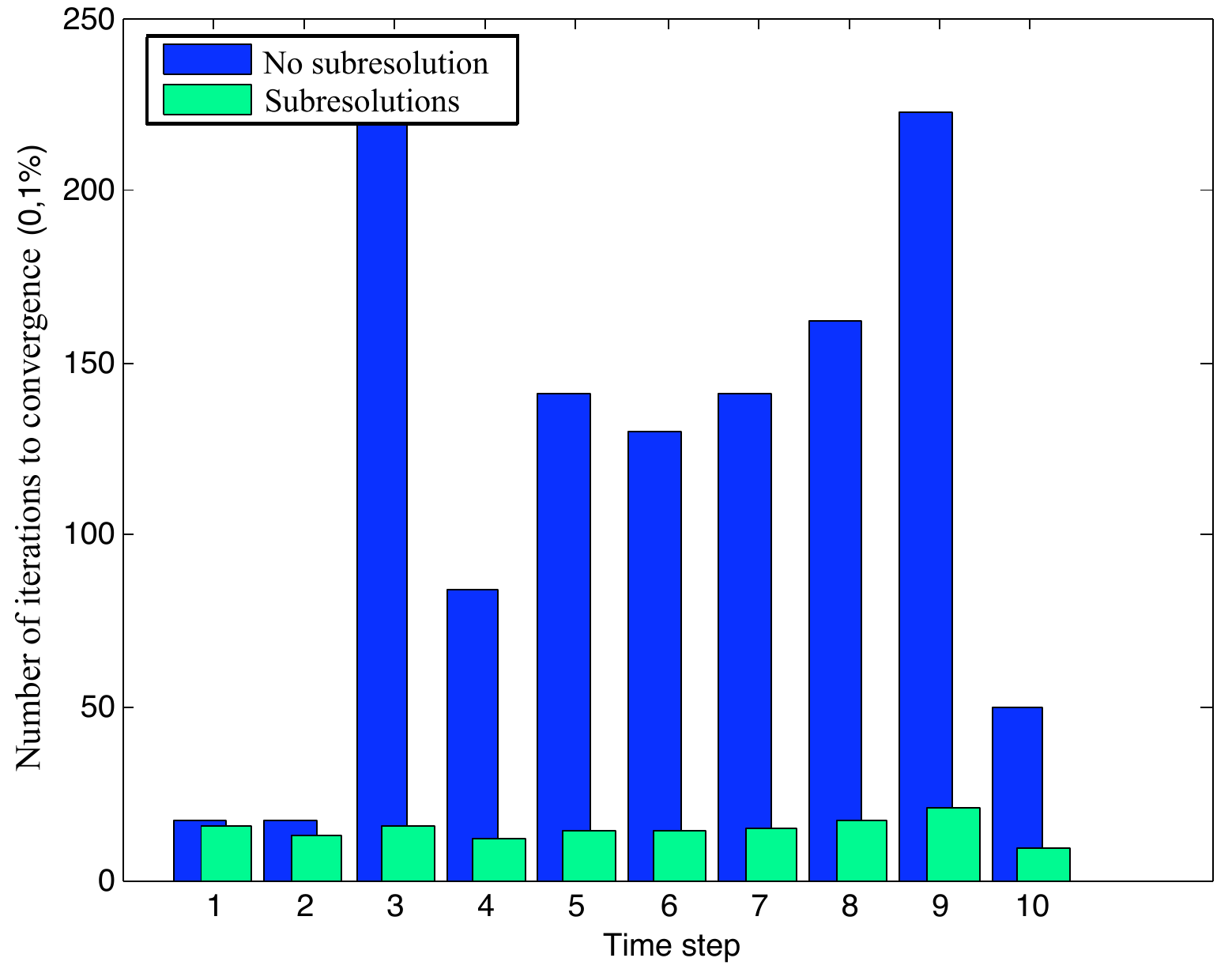}
       \caption{Subresolutions around the tip of the crack: results}
       \label{fig:sub_iterations}
\end{figure}

As it is shown on Figure \ref{fig:sub_iterations}, the resolution of a subproblem around the crack's tip leads to a convergence rate of the global LaTIn solver which is independent on the load increment of the analysis (\textit{i.e.} independent on the area of the interface which becomes delaminated in one increment), which means that the numerical scalability is restored. In addition, the numerical complexity of the strategy is considerably reduced. In our case, the number of local inversions (resolution of problem \eqref{eq:pb_micro}) is divided by two. 

\begin{figure}[htb]
        \centering
        \includegraphics[width=0.9 \linewidth]{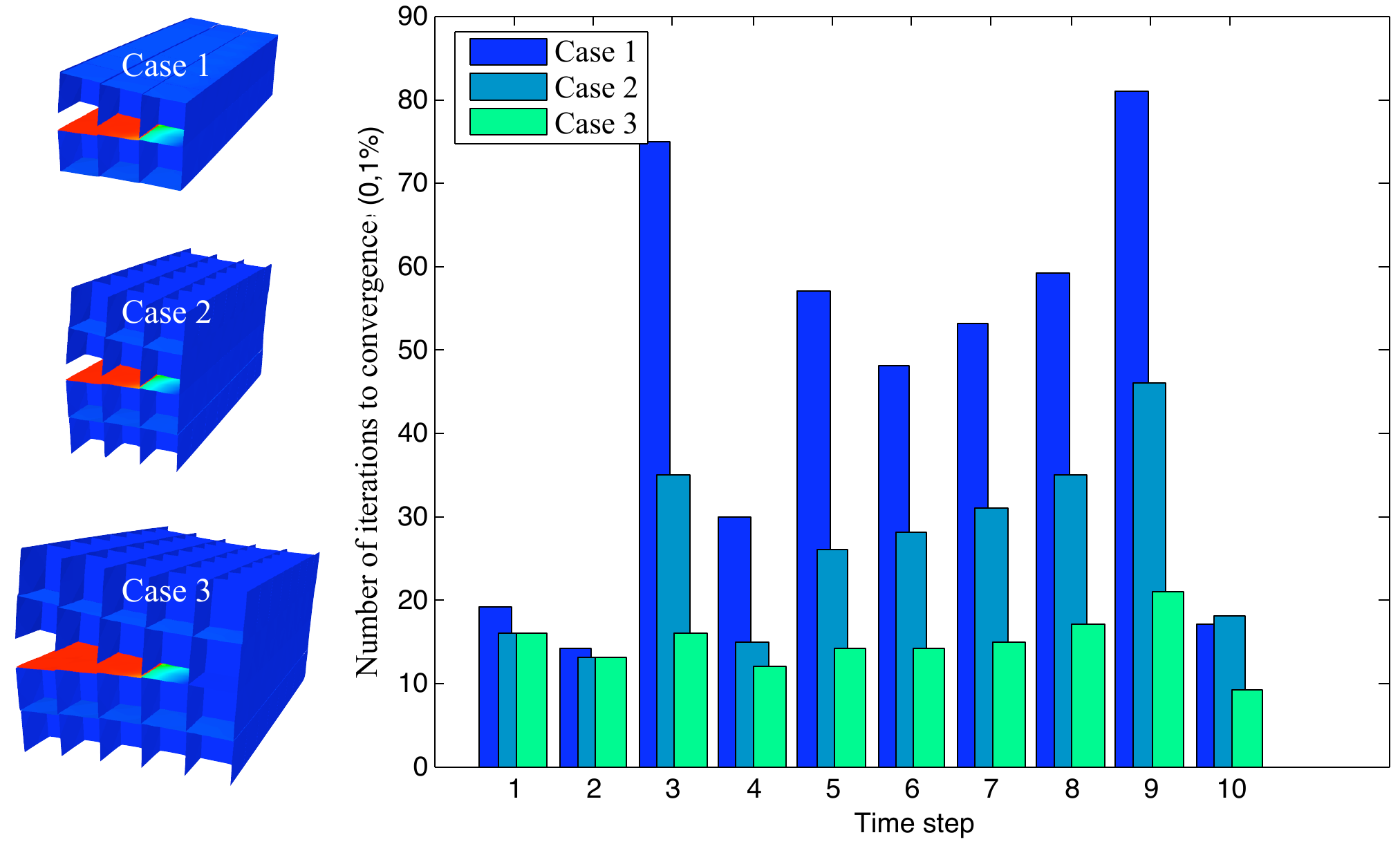}
    \caption{Influence of the size of the subproblem}
    \label{fig:comp_size}
\end{figure}

These results are obtained for a rather ``large'' subresolution box. The influence of the size of the subproblem on the numerical scalability is shown on Figure (\ref{fig:comp_size}). One can clearly see that the number of global LaTIn iterations to convergence drops with an increasing size of the subresolution zone in an asymptotic manner (we observed that no significant gain is observed when increasing the size of the box defined in Case 3). Indeed, as soon as the global effects due to the stress concentration are enclosed in the subresolution zone, no further gain can be obtained by making use of this technique, thus the  optimal size of the subresolution zone seems only to depend on local structural stiffnesses and dimensions (mechanical properties of the process zone) and not on the remaining of the structure.

\section{Discussion}
\label{sec:complex}
The global CPU time is not significantly reduced when implementing this strategy on parallel architectures. Indeed, using the initial allocation among the parallel processors addresses the subresolution to a very small number of processors. The first solution to tackle this difficulty is to perform independent subiterations systematically on all the processors \cite{cresta07,pebrel08}. This strategy has successfully been assessed on various tests though its domain of efficiency is not well mastered yet. 
An other solution is to change the allocation of the substructures among the processors on the fly during the computation  in order to take into account the level of nonlinearity of each set of substructures so that a balanced CPU load is reached.

\begin{figure}[ht]
       \centering
       \includegraphics[width=0.99 \linewidth]{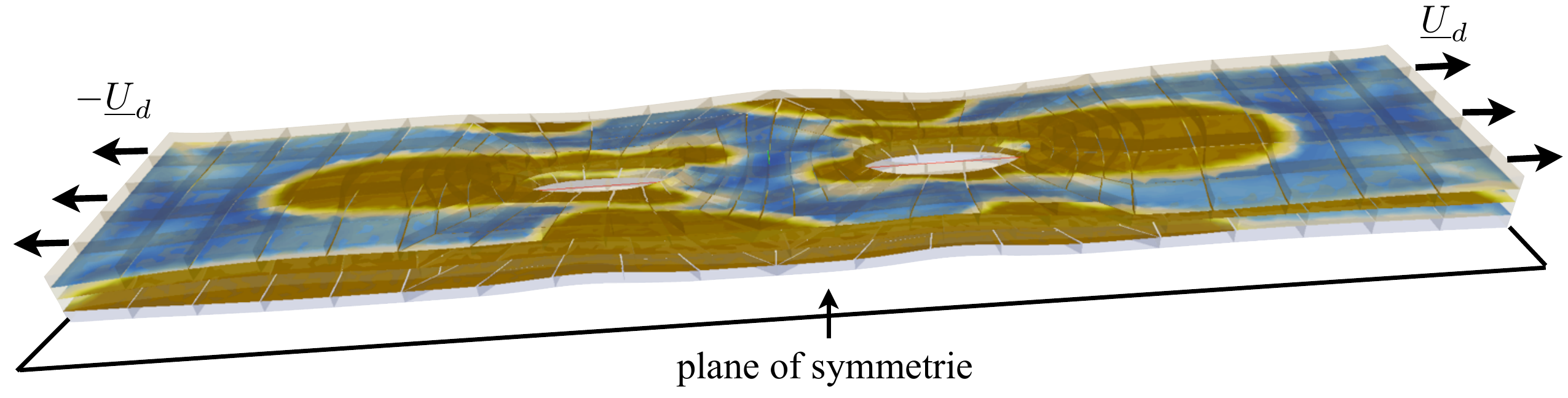}
       \caption{Damage map in the interfaces of a $[0 \ \pm 45 \ 90]_s$ composite holed  plate}
       \label{fig:2_holes}
\end{figure}

The second point to discuss about is the lack of generalization of this dedicated technique.
Figure~\ref{fig:2_holes} clearly illustrates the difficulties that might arise when using the subresolution algorithm in the general case, for multiple crack fronts propagation may be involved (refer to \cite{kerfriden09} for details on the computation of such large debounding problems on parallel computers). The front have an arbitrary shape, which raises the complex issue of the choice of the number and size of the relocalization zones and the problem of potential interactions between relocalization zones. In addition local instabilities might also appear during the relocalization computations, which is not accounted for at this stage of our developments. 

Hence, in the future, the relocalization strategy should be made more general and robust in order to improve the efficiency of the enhanced multiscale domain decomposition strategy for complex laminated structures.

\section{Conclusion}

The accurate prediction of delamination in large process zones of laminated composite structures requires refined models of the material behavior. Such descriptions lead to the resolution of huge systems of equations. In order to compute the exact solution of such a refined model, we used a multiscale domain decomposition strategy based on a LaTIn iterative resolution algorithm. This method is particularly well-suited to solve problems where nonlinearities are introduced in surface joints bonding sets of 3D linear entities.

We have shown that the classical scale separation was insufficient in the high gradient zones to provide numerical scalability. Using an enrichment approach, we proved that a microscopic resolution was required in the process zone at each iteration of the iterative solver. Following this observation, we developed a subresolution procedure which preserved the numerical scalability of the crack propagation parallel calculation. This procedure still needs automation for complex structures (especially concerning the shape and the number of relocalization box(es)) and regulation in case of local instabilities.





\bibliographystyle{plain}
\bibliography{RelocalisationDDM}

\begin{thebibliography}{10}

\bibitem{allix98}
O.~Allix, D.~L\'ev\^eque, and L.~Perret.
\newblock Identification and forecast of delamination in composite laminates by
  an interlaminar interface model.
\newblock {\em Composites Science and Technology}, 58:671--678, 1998.

\bibitem{badea09}
L.~Badea.
\newblock One- and two-level domain decomposition methods for one- and
  two-level domain decomposition methods for nonlinear problems.
\newblock In B.~H.~V. Topping and P.~Iv\'anyi, editors, {\em Proceedings of the
  First International Conference on Parallel, Distributed and Grid Computing
  for Engineering}, 2009.

\bibitem{Cocchetti02}
G.~Cocchetti, G.~Maier, and X.P. Shen.
\newblock Piecewise linear models for interfaces and mixed mode cohesive
  cracks.
\newblock {\em Computer Modelling in Engineering and Sciences}, 3(3):279--298,
  2002.

\bibitem{cresta07}
P.~Cresta, O.~Allix, C.~Rey, and S.~Guinard.
\newblock Nonlinear localization strategies for domain decomposition methods in
  structural mechanics.
\newblock {\em Computer Methods in Applied Mechanics and Engineering},
  196:1436--1446, 2007.

\bibitem{farhat94}
C.~Farhat and F.-X. Roux.
\newblock Implicit parallel processing in structural mechanics.
\newblock {\em Computational Mechanics Advances}, 2:1--24, 1994.

\bibitem{feyel00}
F.~Feyel and J.-L. Chaboche.
\newblock Fe2 multiscale approach for modelling the elastoviscoplastic
  behaviour of long fibre sic/ti composite materials.
\newblock {\em Computer Methods in Applied Mechanics and Engineering},
  183:309--330, 2000.

\bibitem{gendre09}
L.~Gendre, O.~Allix, P.~Gosselet, and F.~Comte.
\newblock Non-intrusive and exact global/local techniques for structural
  problems with local plasticity.
\newblock {\em Computational Mechanics}, 44(2):233--245, 2009.

\bibitem{guidault08}
P.-A. Guidault, O.~Allix, L.~Champaney, and S.~Cornuault.
\newblock A multiscale extended finite element method for crack propagation.
\newblock {\em Computer Methods in Applied Mechanics and Engineering},
  197(5):381--399, January 2008.

\bibitem{Jonsthovel09}
T.B. Jonsthovel, M.B. {Van Gijzen}, C.~Vuik, C.~Kasbergen, and A.~Scarpas.
\newblock Preconditioned conjugate gradient method enhanced by deflation of
  rigid body modes applied to composite materials.
\newblock {\em Computer Modelling in Engineering and Sciences}, 47(2):97--118,
  2009.

\bibitem{kerfriden09}
P.~Kerfriden, O.~Allix, and P.~Gosselet.
\newblock A three-scale domain decomposition method for the 3d analysis of
  debonding in laminates.
\newblock {\em Computational mechanics}, DOI : 10.1007/s00466-009-0378-3, 2009.

\bibitem{ladeveze85}
P.~Ladev\`eze.
\newblock Sur une famille d'algorithmes en m\'ecanique des structures.
\newblock {\em Compte rendu de l'acad\'emie des Sciences}, 300(2):41--44, 1985.

\bibitem{ladeveze99}
P.~Ladev\`eze.
\newblock {\em Nonlinear Computational Structural Mechanics - New Approaches
  and Non-Incremental Methods of Calculation}.
\newblock Springer Verlag, 1999.

\bibitem{ladeveze00}
P.~Ladev{\`e}ze and D.~Dureisseix.
\newblock A micro/macro approch for parallel computing of heterogeneous
  structures.
\newblock {\em International Journal for computational Civil and Structural
  Engineering}, 1:18--28, 2000.

\bibitem{ladeveze06}
P.~Ladev\`eze, G.~Lubineau, and D.~Violeau.
\newblock A computational damage micromodel of laminated composites.
\newblock {\em Int Jal of Fracture}, 137:139--150, 2006.

\bibitem{ladeveze03b}
P.~Ladev{\`e}ze and A.~Nouy.
\newblock On a multiscale computational strategy with time and space
  homogenization for structural mechanics.
\newblock {\em Computer Methods in Applied Mechanics and Engineering},
  192:3061--3087, 2003.

\bibitem{passieux09}
P.~Ladev\`eze, J.C. Passieux, and D.~N\'eron.
\newblock The latin multiscale computational method and the proper generalized
  decomposition (accepted june 2009).
\newblock {\em Computer Methods in Applied Mechanics and Engineering}, 2009.

\bibitem{LETALLEC:1994:NMN}
P.~Le~Tallec.
\newblock {\em Handbook of numerical analysis}, volume~3, chapter Numerical
  methods for nonlinear three-dimensional elasticity, pages 465--622.
\newblock Elsevier science, 1994.

\bibitem{lions90}
P.-L. Lions.
\newblock On the schwartz alternative method. iii: A variant for nonoverlapping
  subdomains.
\newblock In T.F. Chan and al., editors, {\em Third International Symposium on
  Domain Decomposition Methods for Partial Differential Equations}. SIAM, 1990.

\bibitem{lubineau07}
G.~Lubineau, P.~Ladev\`eze, and D.~Marsal.
\newblock Towards a bridge between the micro- and mesomechanics of delamination
  for laminated composites.
\newblock {\em Composites Science and Technology}, 66(6):698--712, 2007.

\bibitem{mandel93}
J.~Mandel.
\newblock Balancing domain decomposition.
\newblock {\em Communications in Numerical Methods in Engineering}, 9(233-241),
  1993.

\bibitem{nouy04}
A.~Nouy and P.~Ladev{\`e}ze.
\newblock On a multiscale computational strategy with time and space
  homogenization for structural mechanics.
\newblock {\em Computer Methods in Applied Mechanics and Engineering},
  192:3061--3087, 2004.

\bibitem{pebrel08}
J.~Pebrel, C.~Rey, and P.~Gosselet.
\newblock A nonlinear dual domain decomposition method: application to
  structural problems with damage.
\newblock {\em Int Jal of Multiscale Computational Engineering}, A paraitre,
  2008.

\bibitem{sanchezpalencia74}
E.~Sanchez-Palencia.
\newblock Comportement local et macroscopique d'un type de milieux physiques
  h{\'e}t{\'e}rog{\`e}nes.
\newblock {\em International Journal of Engineering Science}, 12(4):331--351,
  1974.

\bibitem{potierferry04}
H.~Zahrouni, W.~Aggoune, J.~Brunelot, and M.~Potier-Ferry.
\newblock Asymptotic numerical method for strong nonlinearities.
\newblock {\em Revue Europ\'eenne des Elements Finis}, 13(1-2):97--118, 2004.

\end{thebibliography}

\end{document}